\documentclass[12pt,preprint]{aastex}

\slugcomment{}
 
\newcommand{\msun}{M$_\odot$}

\newcommand{\rosat}{{\it ROSAT}}
\newcommand{\asca}{{\it ASCA}}
\newcommand{\chandra}{{\it Chandra}}
\newcommand{\xmm}{{\it XMM-Newton}}

\shorttitle{Properties of the Brightest Cluster Galaxy and Its Host Cluster}
\shortauthors{Katayama et al.}

\begin{document}

\title{Properties of the Brightest Cluster Galaxy and Its Host Cluster}
\author{Haruyoshi Katayama\altaffilmark{1},  Kiyoshi
Hayashida\altaffilmark{1} and Fumio Takahara\altaffilmark{1}}

\and

\author{Yutaka Fujita\altaffilmark{2}}

\altaffiltext{1}{Department of Earth and Space Science, Graduate School of
Science, Osaka University, 1-1 Machikaneyama, Toyonaka, Osaka 560-0043,
Japan;
hkatayam@ess.sci.osaka-u.ac.jp,hayasida@ess.sci.osaka-u.ac.jp,takahara@vega.ess.sci.osaka-u.ac.jp}
\altaffiltext{2}{National Astronomical Observatory of Japan, 2-21-1,
Osawa Mitaka, Tokyo 181-8588, Japan;yfujita@th.nao.ac.jp}

\begin{abstract}

We investigate the relation between the properties of Brightest Cluster
 Galaxies (BCGs) and those of their host clusters. To quantify the
 properties of cluster hot gas, we employ the parameter $Z$ of the
 fundamental plane of X-ray clusters. It is found that the offset of the
 BCG from the peak of cluster X-ray emission is larger for smaller $Z$
 clusters. The parameter $Z$ (not the redshift {\it z}), which mainly
 depends on virial density $\rho_{\rm {vir}}$, is considered to
 represent the formation epoch of a cluster. We thus consider that the
 offset of the BCG is correlated with the dynamical equilibrium state of
 its host cluster. On the contrary, no significant correlation is found
 between the absolute optical magnitude of the BCG and the parameter
 $Z$.  If the extreme brightness of the BCG is mainly acquired in the
 course of cluster evolution by environmental effect, BCGs are expected
 to be brighter in large $Z$ clusters. Our result is not consistent with
 this simplified view. On the contrary, it is possible that the extreme
 brightness of the BCG is likely to be determined in the early history
 of cluster collapse.

\end{abstract}


\section{Introduction}

Formation and evolution of galaxies are affected by their environment,
particularly when they are embedded in clusters.  We may see such effect
strikingly in brightest cluster galaxies (BCGs), which are defined as
the brightest galaxies among the member galaxies in a cluster. The
luminosities of BCGs are 10 times larger than those of typical normal
field galaxies, and the masses of BCGs are about $10^{13}$ \msun,
comparable to those of galaxy groups. BCGs are not drawn from the same
luminosity function as other cluster member galaxies, suggesting that
they have a distinctive formation history \citep{dressler78}.

According to \citet{dubinski98}, the following three models are proposed
to explain the origin of BCGs: (1) star formation from cooling flows
expected in the centers of clusters \citep{fabian94}; (2) galactic
cannibalism or the accretion of existing galaxies through dynamical
friction and tidal stripping \citep{ostriker77}; and (3) galaxy merging
in the early history of the formation of clusters \citep{merritt85}. In
the cooling flow model, intra-cluster gas gradually condenses at the
high density center of clusters and form BCGs, creating a large
number of new stars.  Therefore, the evolution of BCGs follows that of
their host clusters. We would distinguish the galactic cannibalism model
and the galaxy merging model mentioned above by the formation epoch of
BCGs. A number of small galaxies existing within an evolved cluster form
a BCG in the galactic cannibalism model. On the other hand, in the
galaxy merging model, BCGs are formed during collapse of clusters as
expected in hierarchical cosmological models and have earlier origin
than their host clusters, contrary to the other two models.

The observational properties of BCGs and their host clusters were
studied by \citet{schombert87} and \citet{edge-stewart91}. They found
that the optical luminosity of a BCG is positively correlated with the
X-ray luminosity and hot gas temperature of its host
cluster. \citet{schombert88} observed faint envelopes around 27 cD
galaxies, and found that the luminosity of a cD envelop is correlated
with the X-ray luminosity of its host cluster.  These correlations are
evidence of a close connection between a BCG and its host cluster.  In
this paper, we explore correlations between observational parameters of
BCGs and clusters, using recent X-ray data and focusing on new types of
parameters with particular interest on their evolutionary link. 

We adopt $\Omega_0=1$, $\lambda=0$, and $H_0=50$ km s$^{-1}$
Mpc$^{-1}$ for our study.

\section{Fundamental plane of X-ray clusters}

Besides using observed X-ray parameters of clusters directly, we employ
the parameters defined in the fundamental plane of X-ray clusters
proposed by \citet{fujita99a}. They analyzed the relations among the
central gas density $\rho_{0}$, the core radius $R$, and the temperature
$T$. These data ($\log \rho_{0}$, $\log R$, and $\log T$) lie on a plane
(the fundamental plane) in three dimensional space.  Three axes
($X$,$Y$, and $Z$) of the fundamental plane are written as functions of
$\rho_{0}$ ($10^{-27}$g cm$^{-3}$), $R$ (Mpc), and $T$ (keV). The
equations of these three parameters are
$X=\rho_0^{0.47}R^{0.65}T^{-0.60}$, $Y=\rho_0^{0.39}R^{0.46}T^{0.80}$,
and $Z=\rho_0^{0.79}R^{-0.61}T^{-0.039}$. The scatter of the $X$, $Y$,
and $Z$ are $\Delta \log X = 0.06$, $\Delta \log Y = 0.2$, and $\Delta
\log Z = 0.5$, respectively. Thus, the data distribute on the $Y-Z$
plane.

We especially focus on the parameter $Z$, which is the major axis of
data distribution. \citet{fujita99b} point out that the parameter $Z$ is
regarded as an indicator of cluster age.  According to the virial
equation, the relation between the virial mass $M_{{\rm vir,coll}}$, the virial
radius $R_{{\rm vir,coll}}$, and the virial temperature $T_{{\rm
vir,coll}}$ {\it at the time of cluster collapse} is written as
\begin{equation}
3 k_{{\rm B}} T_{{\rm vir,coll}} = \gamma \mu m_{{\rm H}} \frac{G M_{{\rm vir,coll}}}{R_{{\rm vir,coll}}},
\end{equation} 
where $\mu( = 0.6)$ is the mean molecular weight, $m_{H}$ is the
hydrogen mass, $k_B$ is the Boltzmann constant, $G$ is the gravitational
constant, and $\gamma$ is a fudge factor, which typically ranges between
1 and 1.5. We emphasize that $R_{{\rm vir,coll}}$ is the virial radius
{\it when the cluster collapsed}. Since clusters continue growing,
present $R_{{\rm vir}}$ is different from $R_{{\rm vir,coll}}$. However,
if the structure of the core region is preserved during cluster
evolution, the typical core radius of a cluster $R$ will reflect the
$R_{{\rm vir,coll}}$ as is shown by \citet{salvador98}. Assuming that the gas
temperature $T$ and the core radius $R$ are proportional to $T_{{\rm
vir,coll}}$ and $R_{{\rm vir,coll}}$, respectively, $M_{{\rm vir,coll}}$
is written as $M_{{\rm vir,coll}} \propto R T$.  From $\rho_{{\rm
vir,coll}} \propto M_{{\rm vir,coll}} R_{{\rm vir,coll}}^{-3}$, we
obtain
\begin{equation}
 \rho_{{\rm vir,coll}} \propto R^{-2} T.
\end{equation}
Substituting $R$ and $T$ written as a function of the fundamental
parameters, $\rho_{{\rm vir,coll}}$ is described as
\begin{equation}
 \rho_{{\rm vir,coll}} \propto X^{-1.9} Y^{-0.12} Z^{1.2}.
\end{equation}
Since $X$ has only a small scatter, $\rho_{{\rm vir,coll}}$ mainly
depends on the parameter $Z$. The virial density of a cluster should be
proportional to the critical density of the universe at the epoch when
the cluster collapsed in the spherical collapse model.  This is the
reason why the parameter $Z$ is regarded as an indicator of cluster age
after the collapse.

\section{Offset of the BCG from X-ray Peak}

\begin{figure}[htbp]
\plotone{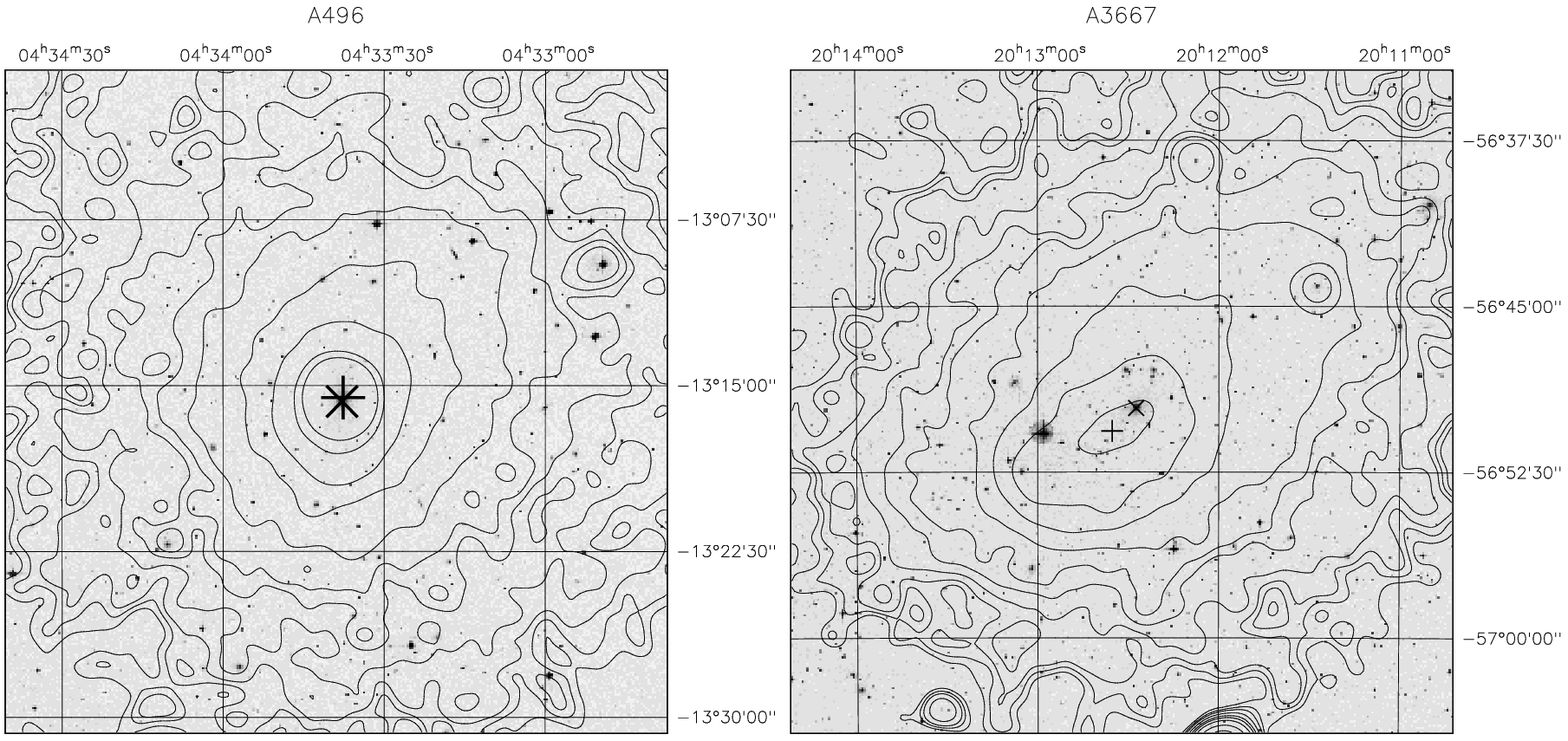} 
\caption{Digitized Sky Survey image of
A496 (left) and A3667 (right). Overlaid contours are \rosat\ PSPC image
in the energy band of 0.1-2.0 keV. ``$\times$'' and ``$+$'' marks
represent the position of the BCG and the X-ray peak, respectively. } 
\label{fig:f1}
\end{figure}

We focus on the position of a BCG in a cluster in addition to the
optical luminosity of a BCG. Fig.\ref{fig:f1} shows the
Digitized Sky Survey image of A496 and A3667. Overlaid contours are
\rosat/PSPC image of 0.1--2.0 keV band. The BCG of A496, which is a
typical cD galaxy, MCG$-$02-12-039, is located at the peak of the X-ray
emission. On the other hand, the BCG of A3667 is located at some
distance from the X-ray peak.  We expect the offset of the BCG from
the peak of the cluster X-ray emission to indicate how the cluster is
close to the dynamical equilibrium state. 

\citet{jones82} proposed a classification scheme of clusters, in which
clusters are categorized by the presence or absence of a cD galaxy which
is giant elliptical galaxy with an extended envelop. Since cD galaxies
are usually located at the center of regular, compact clusters,
significant fraction of BCGs is classified as cD galaxy. Clusters
containing cD galaxies are classified as X-ray dominant (XD); and those
without cD galaxies are classified as non-X-ray dominant (nXD). The
X-ray emissions of XD clusters are strongly peaked on the cD galaxy,
while those of nXD clusters are not associated with any individual
galaxy. \citet{jones82} argues that this classification represents the
state of cluster evolution.

The offset of the BCG from the peak of the cluster X-ray emission might
be regarded as a measure for the XD or nXD categorization. We can
evaluate this quantitatively. Note, however, that BCGs are not
necessarily cDs under our definition. 

\section{Data and Analysis}

We selected 61 $z<0.1$ nearby clusters from the Highest X-ray FLUx Galaxy
Cluster Sample (HIFLUGCS) compiled by \citet{reiprich01}, and 27 $z>0.1$
distant clusters, which are gravitational lensing clusters, from
\citet{hashimotodani99}. 

\citet{reiprich01} collected the gas temperatures from the literature.
\citet{hashimotodani99} determined the gas temperatures from the X-ray
spectra obtained with \asca. Central gas densities and core radii were
determined from the \rosat/PSPC images by \citet{reiprich01} and from
the \rosat/HRI images by \citet{hashimotodani99}.  They fitted surface
brightness profiles with the conventional $\beta$ model,
\begin{equation}
 \rho_{gas}(r) = \frac{\rho_{0}}{(1+(r/R)^2)^{3\beta/2}},
\end{equation}
where $r$ is the distance from the cluster center and $\beta$ is a
fitting parameter. 

We determined the position of the X-ray peak by analyzing \rosat/PSPC
data for nearby cluster sample and \rosat/HRI data for distant cluster
sample.  HRI images were extracted from screened event files by
selecting the pulse height (PI) range of 1--9 ch.  The HRI images were
binned with bin size of 2$''$ and smoothed by a Gaussian kernel of
$\sigma =$ 8--16$''$. The PSPC images were extracted for PI range of
12--200 ch (energy range: 0.1--2.0 keV), binned with 4$''$, and smoothed
by a Gaussian of $\sigma = 16''$.  The correction for telescope
vignetting was not applied to either of the two images.  We define the
central position of the image bin with maximum X-ray intensity as the
position of the X-ray peak.  In some images where bright X-ray point
sources are present, we excluded regions around the sources to explore
the X-ray peak.

Positional uncertainties of the X-ray peak are determined by combining
the error due to counting statistics and telescope pointing accuracy. In
order to estimate the error due to counting statistics, we divided the
original event list for each source into two part and repeated the same
procedure, extraction and peak search, as mentioned above. The
difference between the X-ray peak positions determined from two halves
can be used as an estimate of the error due to counting statistics. The
average statistical errors $\sigma_{stat}$ are $12.7''$ for PSPC and
$13.3''$ for HRI. On the other hand, aspect errors $\sigma_{asp}$ which
are determined by the attitude control of \rosat\ are estimated from the
difference between the position of an X-ray point source in the image,
which is optically identified, and the same obtained from an optical
data base. The average aspect errors are $3.2''$ for PSPC and $2.7''$
for HRI. If we take the positional error as $\sigma =
\sqrt{\sigma_{stat}^{2} + \sigma_{asp}^{2}}$, it corresponds to $\sim$18
kpc at the average redshift of nearby cluster sample and $\sim$75 kpc
at the average redshift of distant cluster sample.

The position of BCGs were taken from the NASA/IPAC Extragalactic
Database (NED)\footnote{\url{http://nedwww.ipac.caltech.edu}}. We
identify the BCG as the galaxy whose optical magnitude is the brightest
among the cluster members. The search radius of a BCG is 15$'$ from the
X-ray peak, which corresponds to 1.2 Mpc at the redshift of
0.05. Because the positional error of the BCGs obtained from NED is
typically about 0.5$''$, the error of the offset of the BCG from the
X-ray peak is dominated by $\sigma_{stat}$. 

In this paper, we also see the correlation of the optical B-band
magnitude of the BCG to other parameters for 26 nearby clusters. These
B-band magnitudes are taken from ``The Third Reference Catalogue of
Bright Galaxies'' by \citet{vaucouleurs91}.

Note that 21 nearby clusters and 5 distant clusters are not included in
our analysis because of the lack of positional information of the
BCG. The number of clusters remained are 40 and 22, for the nearby and
the distant samples, respectively. The average redshifts are 0.05 and
0.3, and the standard deviations of the redshift distribution are 0.02
and 0.1 for each sample.

\section{Results}

We examine the relations between the properties of BCGs and those of
their host clusters.  We employ the parameter $Z$ of the fundamental
plane of X-ray clusters and the hot gas temperature to represent properties
of cluster hot gas.  The offset from the X-ray peak and the optical
absolute magnitude are used to describe the properties of BCGs.

\subsection{Offset of the BCG vs. parameter $Z$}

\begin{figure}[htbp]
 \plotone{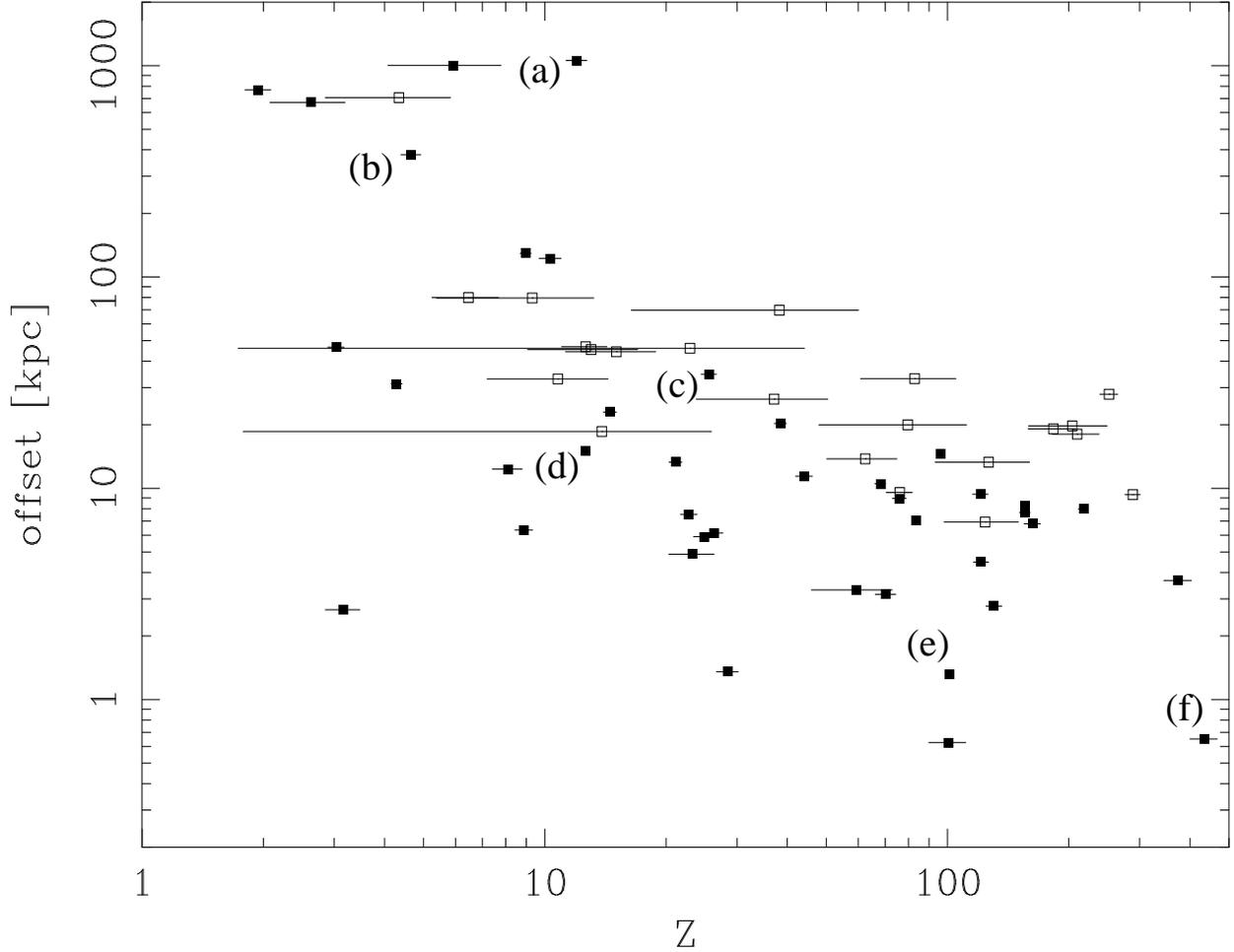} 
\caption{Relation between the parameter $Z$ and the
offset of the BCG. Filled squares and open squares represent the nearby
and the distant clusters, respectively. The typical errors of the offset
are 18 kpc for nearby clusters and 75 kpc for distant clusters.
Individual cluster images marked with (a) to (f) are shown in
Fig. \ref{fig:f4}.}  
\label{fig:f2}
\end{figure}

\begin{figure}[htbp]
 \plotone{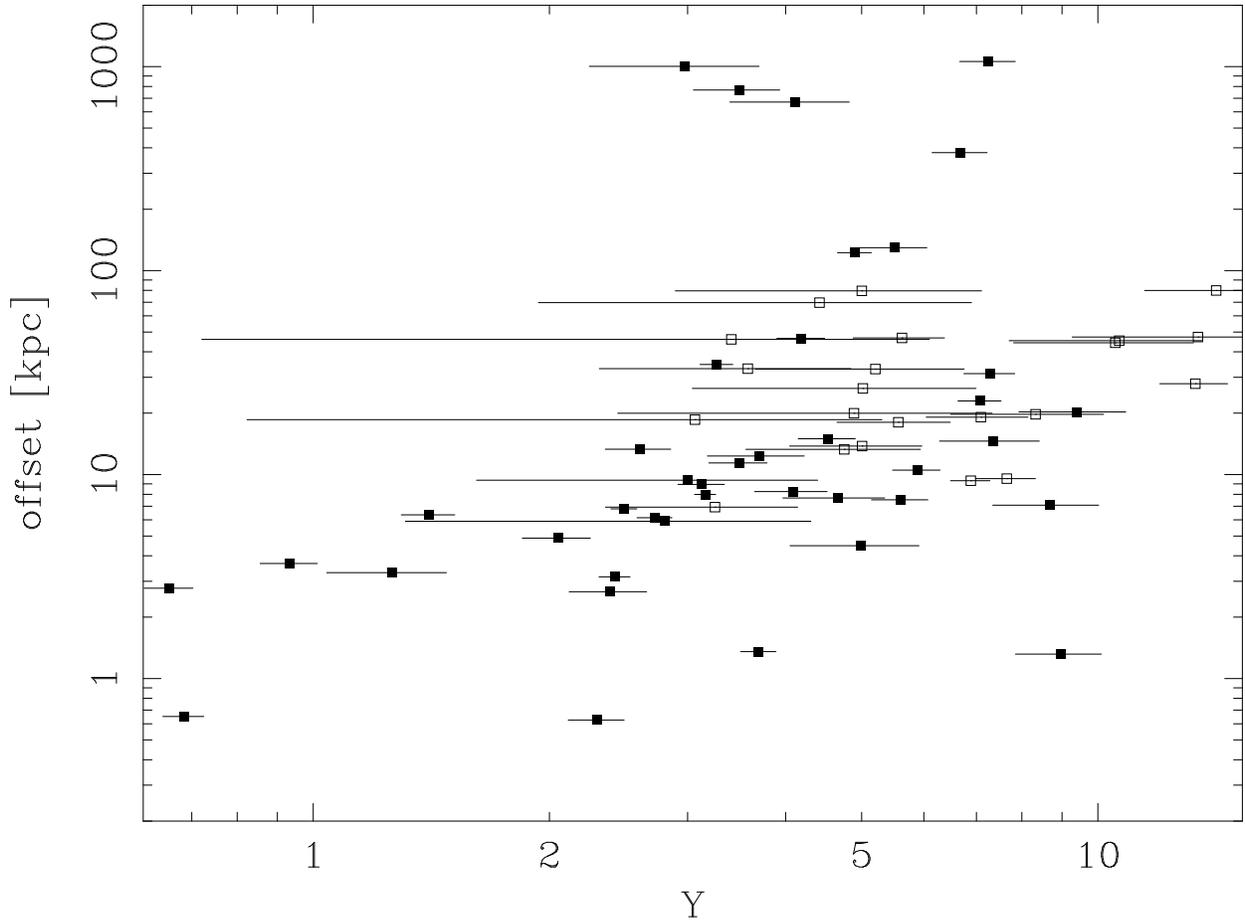}
\caption{Relation between the parameter $Y$ and the offset of
the BCG. Marks are the same in Fig \ref{fig:f2}. }
\label{fig:f3}
\end{figure}

The relations between the offset of the BCG from the X-ray peak and the
parameter $Z$ is shown in Fig.\ref{fig:f2}. Filled squares and open
squares represent nearby clusters and distant clusters, respectively.
The offset of the BCG is larger for the smaller $Z$. The correlation
coefficient is $-0.67$, $-0.74$, and $-0.62$ for nearby, distant, and
all clusters, respectively. On the other hand, the correlation of the
BCG offset to another fundamental plane parameter $Y$ is weaker, as
shown in Fig.\ref{fig:f3}.  The correlation coefficient is 0.27, 0.30
and, 0.33 for each data set, respectively.  As mentioned in section 2,
clusters distribute on the Y-Z plane. These results imply that Z is more
important than Y when we consider the BCG offset. That is the second
reason for us to focus on the parameter Z. We also investigated the
correlation of the BCG offset to the observational quantities
$\rho_{0}$, $R$, and $T$, obtaining the correlation coefficients of
$-0.52$, 0.67, 0.49 for all the data, respectively. Because the
parameters of the fundamental plane are the products of these
quantities, it is not surprising to get similar level of the
correlations to these quantities. It is noted that the hot gas
temperature $T$ depends almost only on the parameter $Y$ ($T \propto
X^{-0.60}Y^{0.80}Z^{-0.039}$), though we will not employ the parameter
$Y$ in the following subsections.

Considering the positional errors of the offset estimated in section 4,
these correlations are not so tight especially for the distant
clusters. However, the offset of the BCG is larger for the smaller $Z$
cluster. In order to test the significance of this trend, we divide the
sample into two groups; one is the large $Z$ clusters ($Z>15$), and the
other is the small $Z$ clusters ($Z<15$). Kolmogorov-Smirnov test shows
the probability that two groups have the same offset distribution are
0.0010, 0.023, and 0.00014 for the nearby, the distant, and all the
clusters, respectively.

Some of the large $Z$ clusters in our sample are cooling flow clusters,
which are well relaxed and often show the central excess emission. We
may have to consider the influence of the central cool component on our
results. \citet{fujita00a} examined the nature of clusters, which have
the central excess considering the inner cool component in addition to
the outer component. They showed that the inner components also satisfy
the same relations of the fundamental plane, though their values of $Z$
are systematically larger than those of the outer components. Therefore,
for some of the large $Z$ clusters in which the cool component is
significant, the value of $Z$ in Fig.\ref{fig:f2} may shift to larger
$Z$ side if we consider this component. On the other hand, if the cool
component reflects a local phenomenon, such as AGN activities, the
ignorance of the component is justified in our study of the global
properties of clusters.

If parameter $Z$ represents the age of the cluster since its formation,
as mentioned in the previous section this correlation indicates the
offset of the BCG is smaller in an aged cluster that spends long
time after its formation.  If the offset of the BCG reflects
deviation from the dynamical equilibrium state of the cluster, as is
often assumed, this might be a natural result.  However, it should be
noted that the parameter $Z$ is derived only from the X-ray properties and
its interpretation by \citet{fujita99a} is a theoretical one. The
negative correlation shown here would be an observational support for
their interpretation of the parameter $Z$ based on galaxy distributions.

In Fig.\ref{fig:f4}, we show some images of individual
clusters marked in Fig.\ref{fig:f2}. The clusters (a)
A754 and (b) A2256 are small Z clusters. These are typical irregular
clusters, and cluster merging might be undergoing.  The cluster (c)
A3562 and (d) A3558 have intermediate $Z$ value.  The X-ray morphology
of these clusters is relatively regular, though some irregularity
remains at their central regions.  The clusters (e) A2029 and (f) NGC5044
are large $Z$ clusters, and have regular morphology even at the central
region, indicating that these cluster are well relaxed system.

We expect that small Z clusters can evolve into large Z clusters through
dynamical relaxation.  The time scale of the evolution is difficult to
evaluate, however, we can provide its rough estimate by comparing the
crossing time of a cluster galaxy and the offset of the BCG
observed. The crossing time $t_{{\rm cr}}$ of a cluster galaxy is
written as
\begin{equation}
 t_{{\rm cr}}(r) \equiv \frac{r}{v_{{\rm r}}} \sim 10^9 ~ {\rm yr} ~ (\frac{r}{{\rm Mpc}}) ~ (\frac{\sigma_{{\rm r}}}{10^3 {\rm km/s}})^{-1},
\end{equation}
where $r$ is the distance from a cluster galaxy to cluster center, $v_r$
is the radial velocity, and $\sigma_{{\rm r}}$ is the radial velocity
dispersion.  Therefore, the evolution time scale of about 1 Gyr is
implied.  Investigation of a large number of distant clusters with
\chandra\ or \xmm\ might enable us to study whether the offset of the BCG
is larger for distant clusters.

\subsection{Optical magnitude of the BCG vs. hot gas temperature}

\begin{figure}[htbp]
 \plotone{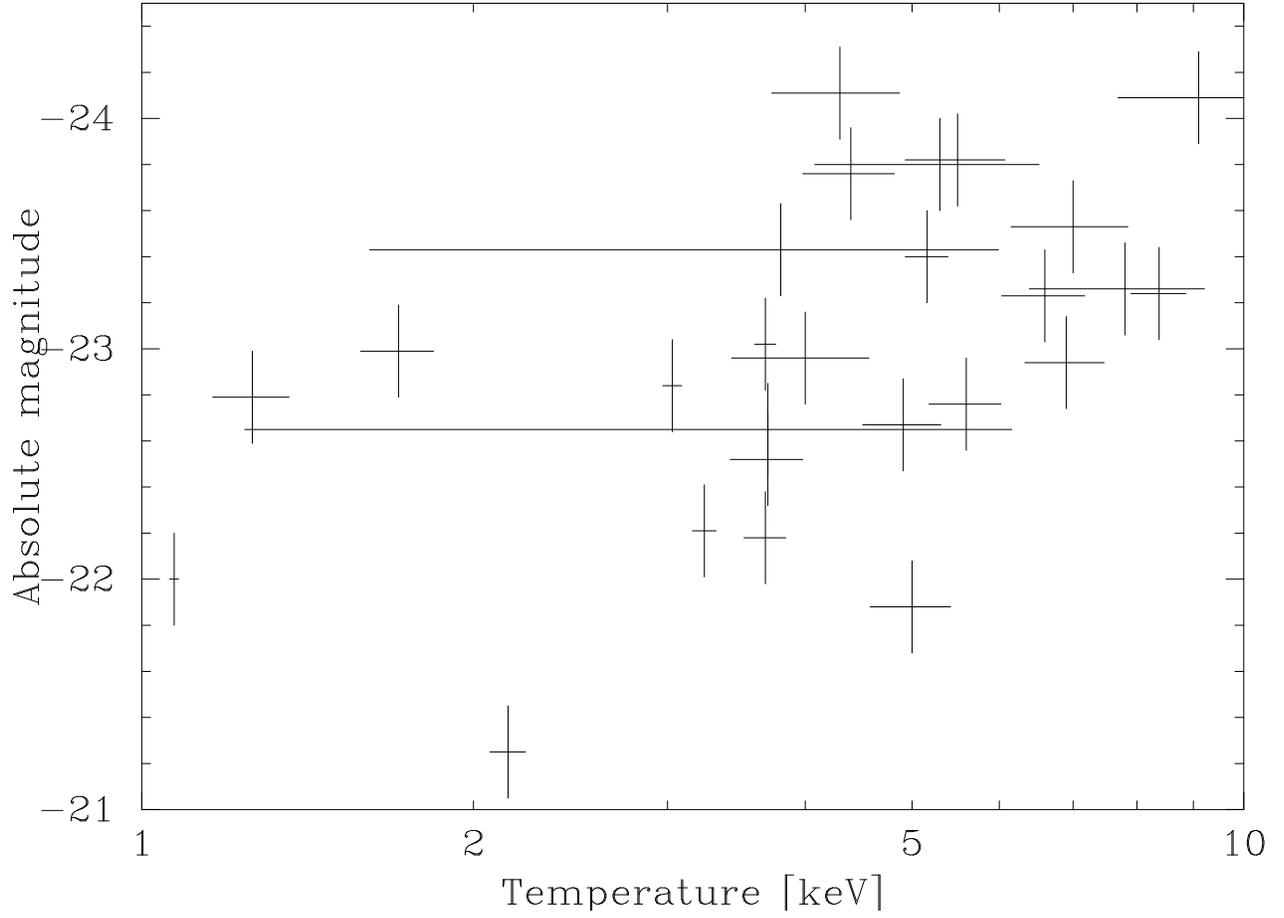} 
\caption{Optical magnitude of the BCG vs. cluster
temperature for the nearby sample. The errors of optical
magnitudes are the typical error of $\pm$ 0.2.} 
\label{fig:f5}
\end{figure}

We next examine the correlation between the hot gas temperature and the
optical magnitude of the BCG. As mentioned in section 4, we restrict the
data of 26 nearby clusters here. As shown in Fig.\ref{fig:f5}, it is found
that the BCGs are brighter in higher gas temperature clusters with a
correlation coefficient of 0.52. A similar correlation was found by
\cite{edge91}, though the scatter of the optical absolute magnitude is 0.35
in their case, smaller than ours (0.62).  

\begin{figure}[htbp]
 \plotone{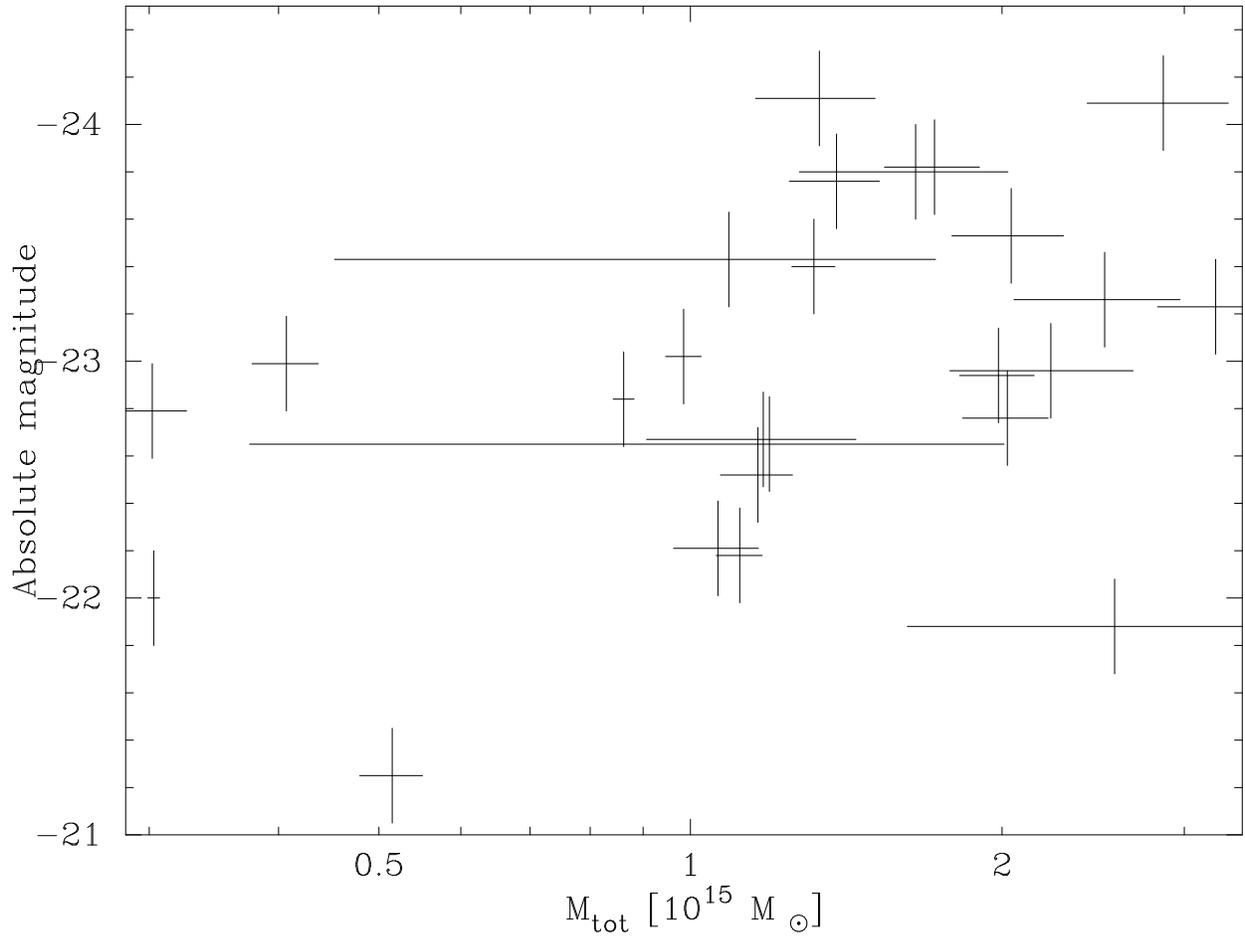} 
\caption{Optical magnitude of the BCG vs. total mass of
 its host cluster.} 
\label{fig:f6}
\end{figure}

Simple arguments based on virial theorem suggest that the mass of a
cluster is simply related to the cluster temperature as $M \propto
T^{3/2}$. Thus, the correlation shown in Fig.\ref{fig:f5} implies that
the optical magnitude of the BCG may also be correlated with the mass of
a cluster. Fig.\ref{fig:f6} shows the relation between the optical
magnitude of the BCG and the total mass of cluster. The total mass is
the integrated mass within the radius of 5 Mpc. Although the correlation is
weaker than that to the temperature (correlation coefficient is 0.43),
larger mass clusters are tend to have brighter BCGs. This suggests that
larger BCGs are formed in the clusters that have deeper potential.

Note that some authors found the similar correlation between the X-ray
luminosity, which is related to the cluster temperature, and the optical
magnitude of the BCG \citep[e.g][]{collins98,burke00}. \citet{brough02},
however, show that this correlation disappears at $z<0.1$, although they
use the K-band magnitude of the BCG.  The B-band magnitude is related to
the star formation in BCGs. The correlation between the B-band magnitude
and the cluster temperature may suggest another possibility that more
stars are formed in the BCG of higher temperature cluster. We need
further investigation for this correlation.

\subsection{Optical magnitude of the BCG vs. parameter $Z$}

\begin{figure}[htbp]
 \plotone{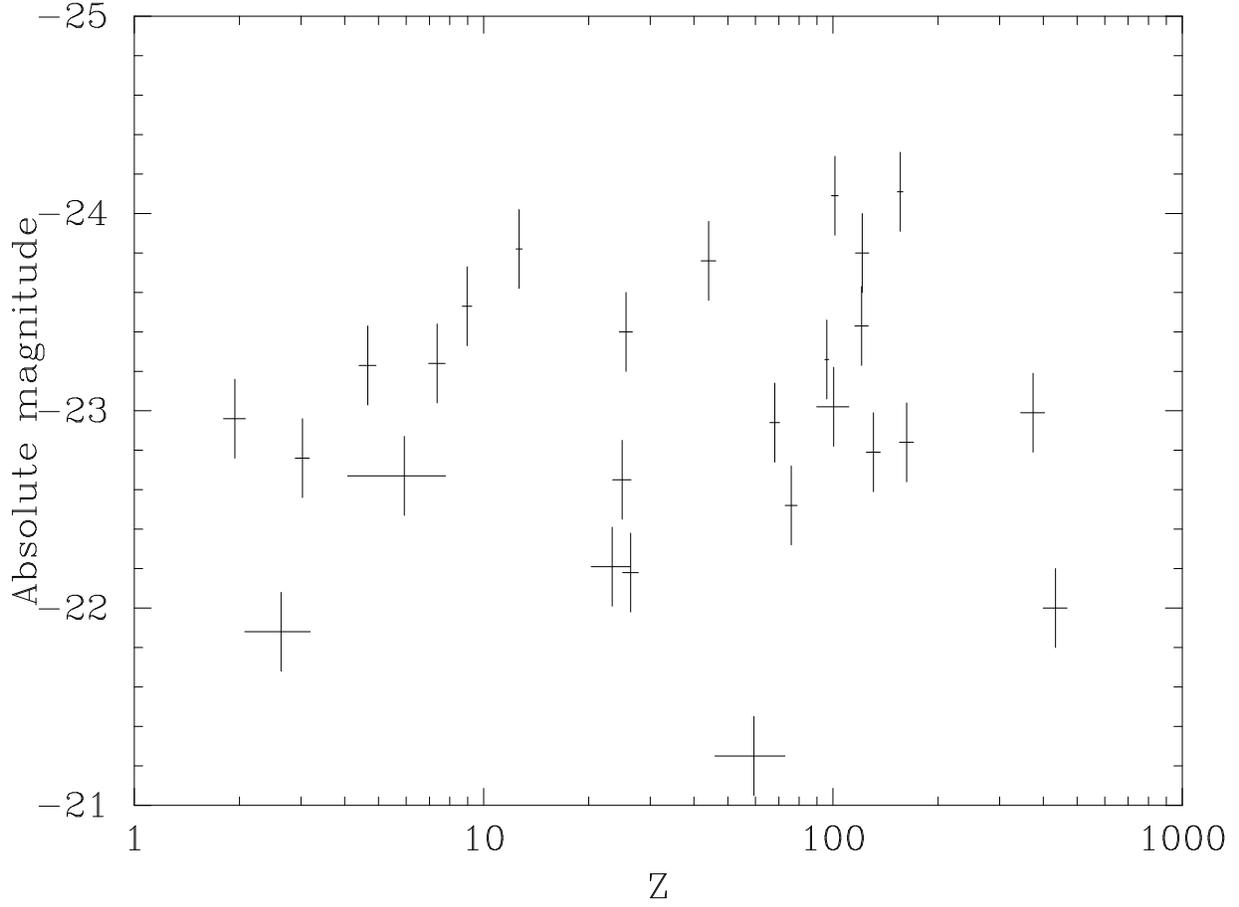} 
\caption{Optical magnitude of the BCG vs. parameter $Z$ of its
host clusters. }
\label{fig:f7}
\end{figure}

In Fig.\ref{fig:f7}, the optical magnitude of the BCG is
plotted against the parameter $Z$. There is not a significant
correlation between these two parameters. The correlation coefficient is
0.10.  The parameter $Z$ is interpreted as an
indicator of cluster age.  Thus, this result suggests that the optical
magnitude of the BCG is not mainly determined by the age of their host
clusters.  Among the three models on the origin of the BCG mentioned in
\citet{dubinski98}, the cooling flow model and the galactic cannibalism
model expect that the extreme luminosity of the BCG is mainly governed
by the evolution of their host clusters. On the other hand, in the
galaxy merging model, the huge BCG luminosity is determined in the
early history of the cluster evolution. The result here is more
favorable to the galaxy merging model than the other two models.

\subsection{Relation between the BCG and the virial density}

\begin{figure}[htbp]
 \plottwo{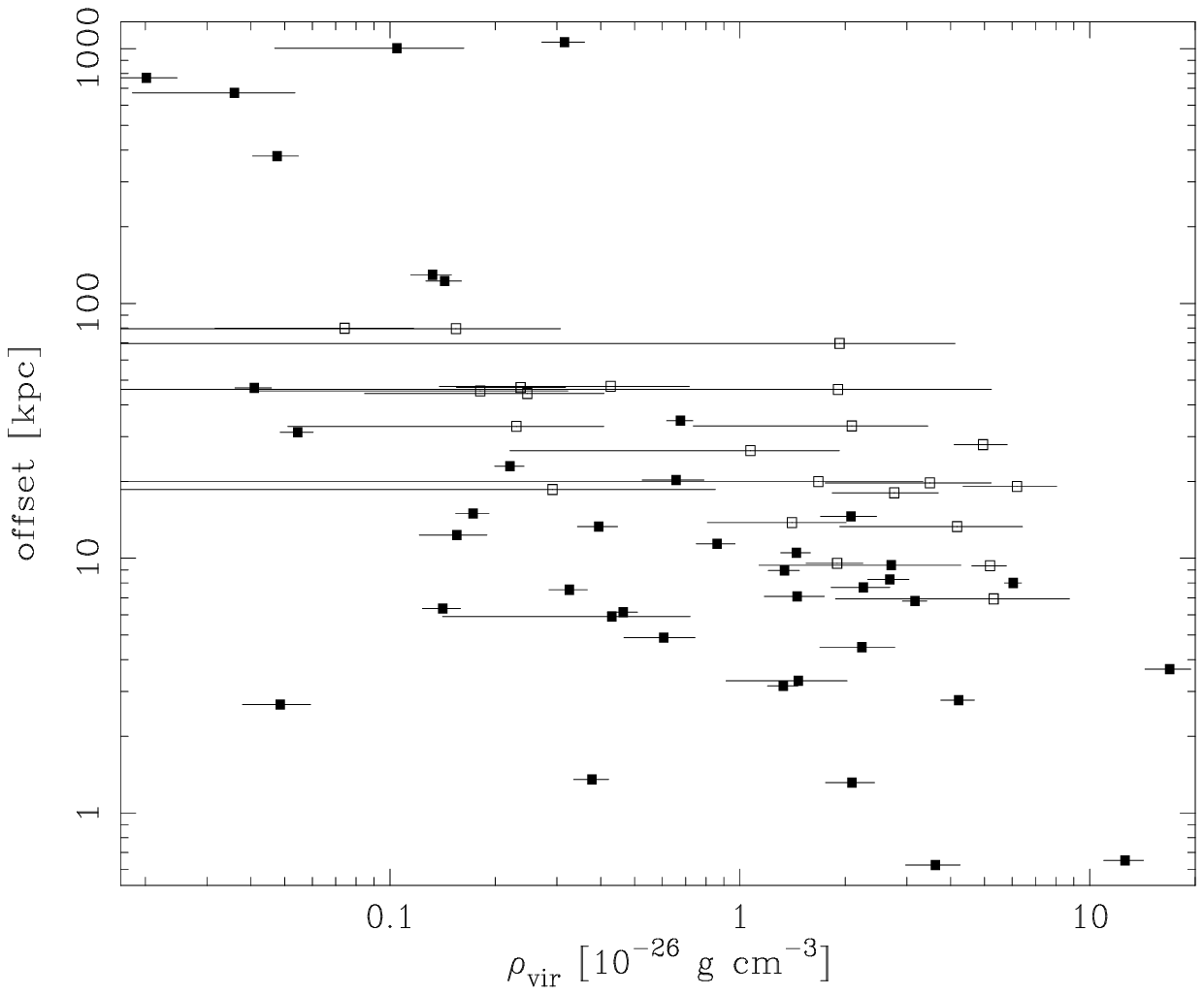}{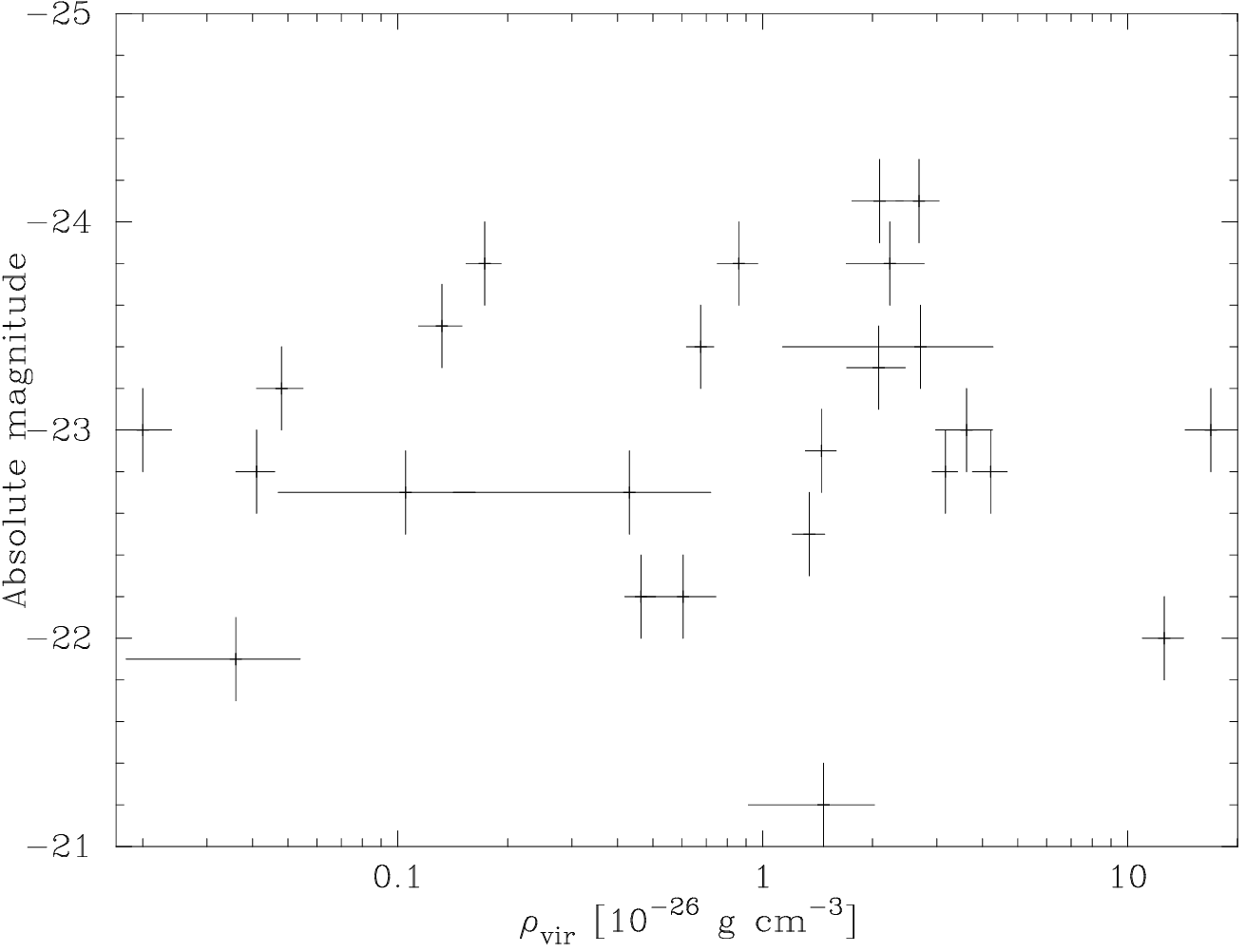} 
\caption{Offset of the BCG vs. virial density (left) and
 optical magnitude of the BCG vs. virial density (right).}
\label{fig:f8}
\end{figure}

As shown in section 2, the parameter $Z$ is closely related to the
virial density of a cluster. We therefore expect similar correlations
when we employ the virial density as a parameter instead of parameter
$Z$. Under the assumption that the virial radius and virial temperature
are proportional to the core radius and gas temperature, we can obtain
the virial density from observable quantities. From equation (1), and
\begin{equation}
 M_{{\rm vir,coll}} = \frac{4 \pi}{3}\rho_{{\rm vir,coll}}R_{{\rm vir,coll}}^{3},
\end{equation}
we obtain
\begin{equation}
 \rho_{{\rm vir,coll}} = \frac{9k_{{\rm B}} T}{4\pi G \mu m_{{\rm H}}}\frac{\beta}{(8R)^{2}},
\end{equation}
where $T_{{\rm vir,coll}} = \gamma \beta T$ and $R_{{\rm vir,coll}} = 8R$ are
assumed following \citet{fujita00b}. In Fig.\ref{fig:f8},
the offset and optical magnitude of the BCG are plotted against the
virial density. The correlation coefficients are -0.54 and 0.06 for the
offset and optical magnitude of the BCG, respectively, confirming
that the virial density acts similarly as the parameter $Z$.

\section{Summary and Discussion}

We have investigated the properties of BCGs and their host clusters. 
The offset of the BCG from the X-ray peak has a negative correlation to
the parameter $Z$ of the fundamental plane of X-ray clusters. Because
the parameter $Z$ is interpreted as an indicator of cluster age since
its collapse, and the offset of the BCG from the cluster center can be
regarded as a good measure of deviation from the dynamical equilibrium
state of a cluster, this correlation tracks the evolution history of the
sample clusters.  Furthermore, since the interpretation of the parameter
$Z$ as the cluster age by \citet{fujita99a} is theoretical one regarding
the hot gas properties of clusters, the correlation to the offset of the
BCG is an observational support to their interpretation.  We also find a
correlation between the cluster temperature and the optical magnitude of
the BCG for nearby clusters, as was found by \citet{edge91}. On the
other hand, there is not a significant correlation among the optical
magnitude of the BCG and the parameter $Z$.

What do these results imply on the origin of BCGs? The offset of
the BCG from the X-ray peak surely reflects deviation from the dynamical
equilibrium state of the cluster and decreases according to the cluster
evolution, which is tracked by the parameter $Z$. On the other hand, the
optical luminosity of the BCG does not have significant correlation to
the parameter $Z$.  These two facts favors the view that the large
luminosity of the BCG is determined prior to or during the early
history of cluster formation.  If the large luminosity of the BCG is
mainly acquired in the course of cluster evolution, we expect some
correlation between the optical luminosity of the BCG and the parameter
$Z$.  Among the three models mentioned in \citet{dubinski98}, therefore,
the galaxy merging model in the early history of cluster formation is
more favorable than the cooling flow model or the galactic
cannibalism model.

The cooling flow model implies the formation of new stars, but there is
a weak evidence for the star formation by cooling flow \citep{mcnamara89}.
Furthermore, recent X-ray observations by \xmm\ showed that the
X-ray emission from cooler gas is much lower than expected from standard
cooling flow models \citep{tamura01,peterson01}. This might imply that
there is a heating mechanism that prevents the ICM from radiative
cooling.  On the other hand, detailed study of the galactic cannibalism
model shows that expected amount of accreted luminosity falls short by an
order of magnitude to account for the BCG luminosity, since the
dynamical friction time scales of galaxies are too long
\citep{merritt85}. As demonstrated by \citet{dubinski98} with his N-body
simulation of a cluster in a hierarchical cosmological model, central
galaxy is formed through the merger of several massive galaxies early in
the cluster's history.  If we consider that the total mass of a cluster
reflects the initial fluctuation of the universe, it may not be strange
that larger BCGs are formed at the bottom of deeper potential of cluster
size. 

Since the significant fractions of BCGs are cD galaxies, the
luminosities of cD galaxies should be also determined prior to or during
the early history of cluster formation. However, \citet{dubinski98}
suggests that the extended envelop of a cD galaxy is not created by the
galaxy merging in the early history of cluster formation. The envelop
might be formed by different process like a tidal stripping.

If BCGs have an earlier origin than their host clusters, we might have a
chance to observe a luminous galaxy which was formed before cluster
collapse and their host clusters are not formed yet. \citet{vikhlinin99}
found X-ray overluminous elliptical galaxies (OLEGs) with \rosat, which
are luminous galaxies comparable to cD galaxies, but have no detectable
galaxies around them. OLEGs thus might be a candidate of such pre-BCG
galaxies. Although extended envelops are not also detected in OLEGs, the
envelop might be formed when an OLEG evolves into a cluster of galaxies.
\\

H.~K. is supported by JSPS Research Fellowship for Young Scientists.
Y.~F. was supported in part by a Grant-in-Aid from the Ministry of
Education, Science, Sports, and Culture of Japan (14740175).

\clearpage

\begin{figure}[htbp]
 \caption{$30'\times30'$ Images of individual clusters marked in
 Fig. \ref{fig:f2}. (a) A754, (b) A2256, (c) A3562, (d)
 A3558, (e) A2029, and (f) NGC5044. Solid lines represent a scale of 200 kpc.}
\label{fig:f4}
\end{figure}

\end{document}